\documentclass[aps,prd,superscriptaddress,tighten]{revtex4}
\usepackage{amsmath}
\usepackage{graphicx}
\usepackage{amsfonts}
\usepackage{amssymb}%
\begin{document}
\title{Semiclassical ultraextremal horizons}
\author{Jerzy Matyjasek}
\email{matyjase@tytan.umcs.lublin.pl, jurek@kft.umcs.lublin.pl}
\affiliation{Institute of Physics, Maria Curie-Sk\l odowska University, pl. Marii
Sk\l odowskiej-Curie 1, 20-031 Lublin, Poland}
\author{O. B. Zaslavskii}
\email{ozaslav@kharkov.ua}
\affiliation{Department of Mechanics and Mathematics, Kharkov V.N. Karazin's National
University, Svoboda Sq.4, Kharkov 61077, Ukraine}
\date{\today}

\begin{abstract}
We examine backreaction of quantum massive fields on multiply-degenerate
(ultraextremal) horizons. It is shown that, under influence of the quantum
backreaction, the horizon of such a kind moves to a new position near which
the metric does not change its asymptotics, so the ultraextremal black holes and
cosmological spacetimes do exist as self-consistent solutions of the semiclassical
field equations.

\end{abstract}
\maketitle






There exist two qualitatively distinct classes of black holes - non-extremal
($T_{H}\neq0$) and extremal ($T_{H}=0$), where $T_{H}$ is the Hawking
temperature. The latter means that the system possesses a degenerate (at least
twice) horizon. In turn, in the latter case the subclass of so-called
ultraextremal horizons (UEH) is singled out. It represents spacetimes with
horizons with triple and higher multiplicity. If one writes the asymptotic
form of the metric coefficient $-g_{00}=f$ $\ $in the Schwarzschild-like
coordinates near the horizon $r_{+}$ as $f\sim(r-r_{+})^{n}$, then $n\geq3$
corresponds to UEH which can be, in principle, black hole or cosmological
horizons. Ultraextremal horizons appear naturally, for instance, in general
relativity and supersymmetrical theories in the spacetimes with the
cosmological constants $\Lambda>0$ and non-zero charge $Q$
(Reissner-Nordstr\"{o}m-de Sitter solution - RNdS), provided some special
relationships between $\Lambda$ and $Q$ hold \cite{rom}. It is worth noting
that recently interest to RNdS revived in the context of dS/CFT correspondence
\cite{mann} and higher-dimensional theories (see. e.g., \cite{Oscar} and
references therein).

Up to now, to the best of our knowledge, spacetimes with ultraextremal
horizons were considered only classically, with quantum backreaction
neglected. Meanwhile, the potential effect of such backreaction on multiple
horizons is not evident in advance since it is unclear whether the condition
of (ultra)extremality is simply slightly shifted or backreaction pushes
classically degenerate horizons away. There was some discussion in literature
on the existence of \textquotedblleft ordinary\textquotedblright\ ($n=2$)
semiclassical extremal black holes (EBH) and it was established that such
solutions do exist \cite{lowe,and+,back}. On the other hand, recent
investigations in two-dimensional dilaton gravity confirmed the existence of
semiclassical EBH but showed that semiclassical UEH are forbidden (except some
very special exactly solvable models) \cite{2dext}.
It is also worth mentioning another issue (closely related to that of EBH) -
quantum-corrected acceleration horizons that arise in the Nariai and
Bertotti-Robinson solutions ~\cite{Od3,Od2,Od1,Kof3,
Kof2,Kof1,Ol2000,Solo,back}.

In the present article we examine the spherically-symmetric UEH metric with
quantum backreaction and show that \textit{self-consistent} solutions
possessing triply degenerated horizon do exist. In doing so, we restrict
ourselves to the case of massive fields in the large mass limit since only in
this case one knows the approximate one-loop stress-energy tensor in terms of
the geometry explicitly (see \cite{bill+,matrn,infeld} and references therein).

The metric under consideration reads
\begin{equation}
ds^{2}=-U\left(  r\right)  dt^{2}+V^{-1}\left(  r\right)  dr^{2}+r^{2}%
d\Omega^{2}\text{,} \label{m}
\end{equation}
and it follows from Einstein equations with the stress-energy tensor $T_{\mu
}^{\nu}$ that%
\begin{equation}
U\left(  r\right)  =e^{2\psi\left(  r\right)  }V\left(  r\right)  \text{,}
\label{uv}
\end{equation}
where%
\begin{equation}
\psi=4\pi\int^{r}drF(r)\text{, \quad}F(r)=r\frac{T_{1}^{1}-T_{0}^{0}}%
{V}\text{.} \label{F}
\end{equation}
As the RNdS-like geometries are not asymptotically flat one can always rescale
time that is equivalent to the change of the integration constant in (\ref{F})
whose particular value is thus unimportant. However, it is unclear in advance
whether $\psi$ remains finite when $r$ approaches UEH (see below).

We assume that the\ right hand side of the Einstein field equations $G_{\mu
}^{\nu}+\Lambda\delta_{\mu}^{\nu}=8\pi T_{\mu}^{\nu}$ is given by $T_{\mu
}^{\nu}=T_{\mu}^{\nu(cl)}+ T_{\mu}^{\nu(q)}$, where the first term
stems from classical source whereas the second one describes the backreaction
of quantum fields on the geometry. As backreaction is considered as a small
perturbation, one could try to use the expansion of the metric taking as the
main approximation the unperturbed metric. However, such a naive approach
suffers from serious shortcomings. It tacitly assumes that parameters of the
classical metric such as the charge, mass, etc. are fixed. Then the metric
which was extremal classically, in general becomes non-extremal (or vice
versa) if quantum corrections are taken into account. (To avoid potential
confusion, this has nothing to do with the third general law since it is not
real transformation of a physical system but, rather, comparison of different
configurations in the space of parameters.) Therefore, we prefer to treat the
problem in a self-consistent way and use the true (quantum-corrected) horizon
value $r_{+}$ from the very beginning. In what follows we restrict ourselves
to the case of the electromagnetic field. Then $T_{1}^{1(cl)}=T_{0}^{0(cl)}$
and it follows from (\ref{uv}), (\ref{F}) that in the zeroth-order of the
approximation (with the backreaction neglected) $U=V$ and we obtain the
classical RNdS metric.

The stress-energy tensor of the quantized massive scalar (with arbitrary
curvature coupling $\xi$), spinor and vector fields \cite{matrn},
\cite{infeld} can be obtained by means of standard methods from the
approximate effective action~\cite{avra1,avra2}
\begin{align}
W_{ren}^{(1)}\,  &  =\,{\frac{1}{192\pi^{2}m^{2}}}\int d^{4}xg^{1/2}\left(
\alpha_{1}^{(s)}R\Box R\,+\,\alpha_{2}^{(s)}R_{\mu\nu}\Box R^{\mu\nu
}\,+\,\alpha_{3}^{(s)}R^{3}\,+\,\alpha_{4}^{(s)}RR_{\mu\nu}R^{\mu\nu}\right.
\nonumber\\
&  +\,\alpha_{5}^{(s)}RR_{\mu\nu\rho\sigma}R^{\mu\nu\rho\sigma}\,+\,\alpha
_{6}^{(s)}R_{\nu}^{\mu}R_{\rho}^{\nu}R_{\mu}^{\rho}\,+\,\alpha_{7}^{(s)}%
R^{\mu\nu}R_{\rho\sigma}R_{~\mu~\nu}^{\rho~\sigma}\nonumber\\
&  +\left.  \,\alpha_{8}^{(s)}R_{\mu\nu}R_{\,\,\,\lambda\rho\sigma}^{\mu
}R^{\nu\lambda\rho\sigma}\,+\,\alpha_{9}^{(s)}{R_{\rho\sigma}}^{\mu\nu}%
{R_{\mu\nu}}^{\lambda\gamma}{R_{\lambda\gamma}}^{\rho\sigma}\,+\,\alpha
_{10}^{(s)}R_{~\mu~\nu}^{\rho~\sigma}R_{~\lambda~\gamma}^{\mu~\nu}%
R_{~\rho~\sigma}^{\lambda~\gamma}\right)  ,
\end{align}
where the spin-dependent numerical coefficients are tabulated in Table I. As
the result of the functional differentiation of $W_{R}$ with respect to the
metric tensor one obtains a rather complicated expression for $T_{\mu}%
^{\nu(q)}$ constructed from the curvature tensor and its covariant
derivatives. To avoid unnecessary proliferation of long formulas we shall not
display it here. An interested reader is referred to\ \cite{matrn},
\cite{infeld} for results and physical motivation.

In the ultraextremal case which we are interested in, the classical metric
functions in (\ref{m}) read \cite{rom}
\begin{equation}
U\left(  r\right)  =V(r)=-\frac{r^{2}}{6\rho^{2}}\left(  1-\frac{\rho}%
{r}\right)  ^{3}\left(  1+\frac{3\rho}{r}\right)  \text{,} \label{u}%
\end{equation}
where $\rho$ is the position of the horizon. It is seen from (\ref{u}) that
the static region is confined by $0<r\leq\rho$, $r=0$ being the singularity.
The quantum state of the field we are dealing with is supposed to be the
Hartle-Hawking one that physically describes the thermal equilibrium, that, in
turn, implies the staticity of the metric in the relevant region. Now, in
contrast to the black hole case the horizon under discussion is of
cosmological nature in the sense that the metric is static for $r<\rho$ and
non-static for $r>\rho$. This does not cause obstacles to the existence of the
Hartle-Hawking state since cosmological horizons are known to possess thermal
properties in a similar way to the black hole case \cite{gh}. However, now
there is a problem connected with the presence of singularity. In the black
hole case (say, for the Schwarzschild metric) the singularity is hidden behind
the horizon, the region in which the metric is static is $r>\rho$. Now the
situation is opposite. To avoid this difficulty, which is not connected by
itself with the issue of UEH, we somewhat modify the system by considering the
following model. Relying on the fact that the singularity in the case under
discussion is time-like, we smear or simply replace it by some central body
with a regular centre and the boundary at $r=R$. Then for $R<r\leq\rho$ we can
use safely the formulas for the stress-energy tensor obtained for the
Hartle-Hawking state.

To evaluate the role of backreaction, we proceed along the same line as in
\cite{back}. We perform two steps: (i) we show that the triple root of $V(r)$
does exist and find corresponding quantum-corrected relationships between
parameters $M$, $Q$, $\Lambda$; (ii) check that the function $F(r)$ is finite,
so that $U(r)$ has the same triple root as $V(r)$. It is convenient to
write\bigskip%

\begin{equation}
V=1-\frac{2m(r)}{r}+\frac{Q^{2}}{r^{2}}-\frac{\Lambda r^{2}}{3}\text{,}
\label{v}%
\end{equation}
where
\begin{equation}
m(r)=M+m_{q}(r)\text{,}%
\end{equation}%
\begin{equation}
M=\frac{r_{+}}{2}+\frac{Q^{2}}{r_{+}}-\frac{\Lambda r_{+}^{3}}{6}\text{.}%
\end{equation}
$m_{q}=-4\pi\int_{r_{+}}^{r}dr^{\prime}r^{\prime2}T_{0}^{0(q)}$ is the
contribution of quantum fields, $m(r_{+})=0$. To get rid off the denominators
in (\ref{v}), it is convenient to introduce $g(r)\equiv r^{2}V(r)$.

In the vicinity of the horizon
\begin{equation}
m(r)=M+A(r-r_{+})+...\text{,}\label{mA}%
\end{equation}
where $A=-4\pi T_{0}^{0(q)}r_{+}^{2}$ is a small parameter responsible for
backreaction. We should check that equations
\begin{equation}
g(r_{+})=0=g^{\prime}(r_{+})=g^{\prime\prime}(r_{+})\label{g}%
\end{equation}
are self-consistent. The form (\ref{mA}) is unappropriate for analysis of the
aforementioned equations since the equations would contain as the parameter
their own root that, as is shown in~\cite{back} leads to difficulties
connected with the appearance of spurious roots \cite{lowe}. To avoid this
difficulty, we redefine $M_{0}=M-Ar_{+}$ and substitute in (\ref{v}) the
expression $m(r)=M_{0}+Ar$. Then it is straightforward to show that all three
equations (\ref{g}) are mutually consistent, with
\begin{equation}
r_{+}^{2}=\frac{1}{2\Lambda}(1-2A)\text{,}%
\end{equation}%
\begin{equation}
M_{0}=\frac{1}{\sqrt{2\Lambda}}(1-3A)\text{,}%
\end{equation}%
\begin{equation}
Q^{2}=\frac{1}{4\Lambda}\left(  1-4A\right)  .\text{ }%
\end{equation}
For the massive scalar, spinor and vector fields considered in this paper one
has respectively
\begin{equation}
A^{\left(  0\right)  }=\frac{\Lambda^{2}\left(  3780\eta^{3}+63\eta-8\right)
}{5670\pi m^{2}},\qquad A^{\left(  1/2\right)  }=-\frac{\Lambda^{2}}{252\pi
m^{2}},\qquad A^{\left(  1\right)  }=-\frac{2\Lambda^{2}}{105\pi m^{2}%
},\label{A}%
\end{equation}
where superscripts denote the value of spin and $\eta=\xi-1/6$ and the form
(\ref{A}) implies that the parameters $M$, $Q$, $\Lambda$ are not arbitrary
but connected by the relationship inherent to the ultraextremal case.

Now we pass to the next step and substitute into the expression for $T_{\mu
}^{\nu(q)}$ the metric (\ref{m}) in which we put, in the main approximation,
$U=V=V_{0}$. \ The function $\psi\left(  r\right)  $ is given by Eq.~(\ref{F})
with
\begin{equation}
\text{\quad}F(r)=r\frac{T_{1}^{1\left(  q\right)  }-T_{0}^{0\left(  q\right)
}}{V_{0}}\text{.}\label{11}%
\end{equation}
If $\psi(r_{+})$ is bounded, the backreaction does not change qualitatively
the character of the metric. It is worth noting that the finiteness of
$\psi(r_{+})$ is equivalent of the finiteness of the energy measured by an
observer who moves along the radial geodesics \cite{lha}. Specifically, for
the line element (\ref{u}) one obtains after calculations
\begin{equation}
\psi_{1}\left(  r\right)  =\frac{1}{\pi m^{2}}\sum_{i=4}^{8}\Lambda^{-\left(
i-4\right)  /2}B_{i}^{\left(  s\right)  }r^{-i}+C_{1}^{\left(  s\right)  },
\end{equation}
where $B_{i}^{\left(  s\right)  }$ with $B_{5}^{\left(  s\right)  }=0$ are
coefficients depending on the parameters of the theory and $C_{i}^{\left(
s\right)  }$ are integration constants. Specifically, for the massive
$s=0,\,1/2,\,1$ fields one has
\begin{equation}
B_{4}^{\left(  s\right)  }=-h\left(  0\right)  \left(  \frac{\eta}{120}%
-\frac{1}{4320}\right)  -\frac{h\left(  1/2\right)  }{2880}-h\left(
1\right)  \frac{23}{4320},
\end{equation}%
\begin{equation}
B_{6}^{\left(  s\right)  }=h\left(  0\right)  \left(  \frac{14}{135}\eta
-\frac{7}{3240}\right)  +h\left(  1/2\right)  \frac{13}{1080}+h\left(
1\right)  \frac{211}{1080},
\end{equation}%
\begin{equation}
B_{7}^{\left(  s\right)  }=-2^{1/2}\left[  h\left(  0\right)  \left(  \frac
{4}{45}\eta-\frac{17}{6615}\right)  +h\left(  1/2\right)  \frac{13}%
{1470}+h\left(  1\right)  \frac{1223}{6615}\right]  ,
\end{equation}%
\begin{equation}
B_{8}^{\left(  s\right)  }=h\left(  0\right)  \left(  \frac{13}{320}%
\eta-\frac{47}{26880}\right)  +h\left(  1/2)\right)  \frac{37}{17920}%
+h\left(  1\right)  \frac{2141}{26880},
\end{equation}
where $h\left(  s\right)  $ is the number of fields with spin $s$. We see that
$\psi(r_{+})$ is bounded and this key feature entails immediately the
conclusion that semiclassical UEH of the type under discussion do exist.

To summarize, we showed that there exist self-consistent UEH solutions of
semiclassical field equations, if the corresponding classical system admits
them. Although we restricted ourselves by one concrete physically relevant
example (RNdS metric) the general approach applies to a more general situation
when the function $f$ has near the asymptotics \thinspace$(r-r_{+})^{n}$,
$n\geq3$. In this case it also turns out that the functions $F(r)$ and
$\psi(r)$ are finite near the horizon, so that $U(r)$ has the root of same
multiplicity $n$. In this sense, ultraextremal horizons are stable against
backreaction of massive fields. In particular, this confirms the significance
of ultraextremal black holes in addition to ordinary EBH as potential
candidates to the role of stable remnants after black hole evaporation.


\begin{table}
\caption{The coefficients $\alpha_{i}^{(s)}$ for the massive scalar, spinor,
and vector field}%
\label{table1}
\begin{ruledtabular}
\begin{tabular}
[b]{cccc}
& s = 0 & s = 1/2 & s = 1\\
$\alpha^{(s)}_{1} $ & ${\frac{1}{2}}\xi^{2}\,-\,{\frac{1}{5}} \xi$%
\,+\,${\frac{1}{56}}$ & $- {\frac{3}{140}}$ & $- {\frac{27}{280}}$\\
$\alpha^{(s)}_{2}$ & ${\frac{1}{140}}$ & ${\frac{1}{14}}$ & ${\frac{9 }{28}}%
$\\
$\alpha^{(s)}_{3}$ & $\left(  {\frac{1}{6}} - \xi\right)  ^{3}$ & ${\frac
{1}{432}}$ & $- {\frac{5}{72}}$\\
$\alpha^{(s)}_{4}$ & $- {\frac{1}{30}}\left(  {\frac{1}{6}} - \xi\right)  $ &
$- {\frac{1}{90}}$ & ${\frac{31}{60}}$\\
$\alpha^{(s)}_{5}$ & ${\frac{1}{30}}\left(  {\frac{1}{6}} - \xi\right)  $ & $-
{\frac{7}{720}}$ & $- {\frac{1}{10}}$\\
$\alpha^{(s)}_{6}$ & $- {\frac{8}{945}} $ & $- {\frac{25 }{376}}$ & $-
{\frac{52}{63}}$\\
$\alpha^{(s)}_{7}$ & ${\frac{2 }{315}}$ & ${\frac{47}{630}}$ & $- {\frac
{19}{105}} $\\
$\alpha^{(s)}_{8}$ & ${\frac{1}{1260}}$ & ${\frac{19}{630}} $ & ${\frac
{61}{140}} $\\
$\alpha^{(s)}_{9}$ & ${\frac{17}{7560}}$ & ${\frac{29}{3780}}$ & $- {\frac
{67}{2520}}$\\
$\alpha^{(s)}_{10}$ & $- {\frac{1}{270}}$ & $- {\frac{1}{54}} $ & ${\frac
{1}{18}}$%
\end{tabular}
\end{ruledtabular}
\end{table}


\end{document}